\documentclass[10pt,conference]{IEEEtran}
\IEEEoverridecommandlockouts

\usepackage{cite}
\usepackage{amsmath,amssymb,amsfonts}
\usepackage{algorithmic}
\usepackage{graphicx}
\usepackage{textcomp}
\usepackage{xcolor}
\usepackage{url}
\usepackage[utf8]{inputenc}
\usepackage{array}
\usepackage{listings}
\usepackage{balance}
\usepackage{longtable,booktabs}
\usepackage{todonotes}
\usepackage{tabularx}
\usepackage{pifont}
\usepackage[inline]{enumitem}
\usepackage{orcidlink}
\usepackage{tailwindcolors}
\usepackage{fontawesome}
\usepackage[framemethod=TikZ]{mdframed}
\usepackage{dirtytalk}
\usepackage{hyperref}
\usepackage{comment}
\usepackage{threeparttable}
\hypersetup{hidelinks}

\ifCLASSOPTIONcompsoc
\usepackage[caption=false,font=normalsize,labelfont=sf,textfont=sf]{subfig}
\else
\usepackage[caption=false,font=footnotesize]{subfig}
\fi

\def\BibTeX{{\rm B\kern-.05em{\sc i\kern-.025em b}\kern-.08em
    T\kern-.1667em\lower.7ex\hbox{E}\kern-.125emX}}

\definecolor{mairieli}{HTML}{ff75ad}

\definecolor{kiev}{HTML}{f5b027}

\definecolor{box_bg}{named}{purple100}
\definecolor{box_line}{named}{purple300}
\definecolor{box_font}{named}{purple900}

\definecolor{experience}{named}{violet800}
\definecolor{impact}{named}{red800}
\definecolor{regulation}{named}{blue800}
\definecolor{participation}{named}{green800}
\definecolor{context}{named}{yellow800}

\mdfdefinestyle{textbox}{%
linewidth=3pt,%
leftmargin=0cm,%
rightmargin=0cm,%
topline=false,%
rightline=false,%
bottomline=false,%
leftline=true,%
skipabove=0pt,%
skipbelow=0pt,%
}

\newcommand{\refRQBaseColor}[1]{%
    \hyperref[RQ:#1]{\textbf{RQ\textsubscript{#1}}}%
}%

\definecolor{rqpurple}{HTML}{7A61DD}
\definecolor{rqbackground}{HTML}{F3EEF9}

\usepackage{tcolorbox}
\newtcolorbox{rqbox}{
    colback=rqbackground,
    colframe=rqpurple,
    boxrule=0pt,
    leftrule=2pt,
    arc=0pt,
    boxsep=0pt,
    left=7pt,
    right=7pt,
    top=5pt,
    bottom=5pt,
    before skip=8pt,
    after skip=5pt
}

\newcommand{\researchquestion}[2]{%
    \begin{rqbox}
        \textbf{\textcolor{rqpurple}{#1}\enspace #2}
    \end{rqbox}%
}

\begin{document}

\title{``\textit{My Code Runs Fine, I Don't}'': Understanding How Software Professionals Work Through Chronic Pain}

%\author{\IEEEauthorblockN{Anonymous Author(s)}
%IEEEauthorblockA{
%\textit{Anonymous University}\\
%Anonymous Country}
%}

% \author{
% \IEEEauthorblockN{Mairieli Wessel}
% \IEEEauthorblockA{\textit{Radboud University} \\
% The Netherlands} \and

% \IEEEauthorblockN{Kiev Gama}
% \IEEEauthorblockA{\textit{Federal University of Pernambuco} \\
% Brazil} \and

% \IEEEauthorblockN{Joukje Oosterman}
% \IEEEauthorblockA{\textit{Radboud University} \\
% The Netherlands} \and

% \IEEEauthorblockN{Ronnie de Souza Santos}
% \IEEEauthorblockA{\textit{University of Calgary} \\
% Canada}
% }

\author{%
\IEEEauthorblockN{%
Mairieli Wessel\IEEEauthorrefmark{1},
Kiev Gama\IEEEauthorrefmark{2},
Joukje Oosterman\IEEEauthorrefmark{1}\IEEEauthorrefmark{3}, and
Ronnie de Souza Santos\IEEEauthorrefmark{4}
}
\IEEEauthorblockA{%
\IEEEauthorrefmark{1}Radboud University, the Netherlands\\
\IEEEauthorrefmark{2}Federal University of Pernambuco, Brazil\\
\IEEEauthorrefmark{3}Donders Institute for Brain, Cognition and Behavior, the Netherlands\\
\IEEEauthorrefmark{4}University of Calgary, Canada
}
}

\maketitle

\begin{abstract}
Software development activities rely on sustained concentration, prolonged computer use, and collaboration. The effort and personal costs of meeting these demands are not always visible to others. Despite growing interest in developer well-being, stress, and mental health, chronic pain remains largely overlooked. We investigated how software professionals experience and navigate chronic pain at work by interviewing 34 professionals living with chronic pain. Our findings highlight the continuous, often invisible effort involved in managing pain while continuing to contribute. Participants reported that prolonged static posture, unsuitable workstations, intensive keyboard and mouse use, workload pressure, and deep concentration aggravated their pain. Participants described physical, cognitive, and emotional experiences that affected task completion, perceived technical quality, and collaboration. Adapting tools, workspaces, routines, and schedules helped them remain engaged, but continued participation often involved recovery debt, persistent symptoms, and concerns about career sustainability. Our main contribution is a conceptual framework of chronic pain in software development, which we discuss in relation to the biopsychosocial model of pain. This framework may help practitioners and companies identify demands that may aggravate pain and consider adjustments, while guiding research on more inclusive and sustainable software engineering workplaces.
\end{abstract}


\section{Introduction}

Software development is a cognitively demanding and collaborative activity. It involves reasoning about complex systems, writing and reviewing code, and debugging failures~\cite{latoza2006maintaining,meyer2014software}. It also requires coordination with teammates to refine requirements, manage dependencies, and meet delivery schedules~\cite{kula2022factors}. Mental fatigue has been reported to make demanding programming tasks more difficult~\cite{sarkar2017characterizing}. Sleep deprivation has also been associated with poorer programming performance among novice developers~\cite{fucci2018need}. Alongside these individual factors, work environment, team culture, and other social and technical conditions have been associated with job satisfaction and perceived productivity~\cite{storey2019towards}.

Chronic pain adds another consideration to understanding participation at work. It is a long-term condition that persists or recurs for months or years~\cite{raja2020revised,nicholas2019iasp} and has been associated with presenteeism, absenteeism, reduced productivity, disability, and early retirement~\cite{kawai2017adverse,patel2012impact,blake2025employees,mcdonald2011musculoskeletal}. The effects of chronic pain are often invisible to others. Someone may appear fully engaged in a meeting, respond to messages, or write code while managing pain, fatigue, difficulty concentrating, medication side effects, or the risk of a \textit{flare-up} (a temporary worsening of pain beyond its usual level). Others may see someone participating at work without recognizing the effort it takes to do so. Understanding these experiences is important for designing workplaces, tools, and practices that support sustainable participation.

Software engineering research has increasingly examined well-being, stress, burnout, and developer experience~\cite{martinez2026factors,arriel2026understanding,wong2023mental,graziotin2018happens,fagerholm2012developer,greiler2022actionable}, but chronic pain has received little attention. This omission is relevant because developer experience research recognizes cognitive and affective dimensions of development, while chronic pain has been associated with difficulties in attention, memory, and executive function~\cite{higgins2018relationship,zhang2021evidence,moore2012interruptive}. Pain and occupational health research has examined general working populations and computer users, but has rarely considered software professionals as a distinct occupational group. Less is therefore known about how chronic pain interacts with the sustained reasoning, use of development tools, and team coordination involved in software development.

We investigate this gap through an exploratory interview study with 34 software professionals living with chronic pain. A pre-interview questionnaire collected demographic, professional, and standardized pain-profile information to contextualize semi-structured interviews about their experiences and pain management practices. Our study addresses the following research question:

\researchquestion{RQ.}{How do software professionals experience and navigate chronic pain at work?}

Participants reported that prolonged static posture, poor ergonomic fit, intensive keyboard and mouse use, concentration demands, workload pressure, and conditions associated with remote work and travel aggravated their pain. They described physical, cognitive, and emotional experiences that affected work capacity, task completion, perceived technical quality, task selection, and collaboration. To manage symptoms and remain engaged, they adapted their routines, work practices, tools, environments, and schedules. Continued participation often involved \textit{recovery debt} (an accumulated need for rest when recovery was postponed to keep working), persistent symptoms, and concerns about long-term career sustainability.

This paper makes three contributions: (1) an empirically grounded conceptual framework showing how chronic pain interacts with software development activities, including programming and debugging, typing, task selection, perceived technical quality, and collaboration; (2) a discussion positioning these findings within the biopsychosocial model of chronic pain~\cite{gatchel2007biopsychosocial} and motivating future investigation of a physical dimension of Developer Experience (DX)~\cite{fagerholm2012developer,greiler2022actionable}; and (3) practical implications grounded in participants' reported adaptations, alongside literature-informed proposals for future evaluation.

\section{Background and Related Work}

\subsection{Chronic Pain}
The International Association for the Study of Pain (IASP) defines \textbf{pain} as \say{an unpleasant sensory and emotional experience associated with, or resembling that associated with, actual or potential tissue damage}~\cite{raja2020revised}. More specifically, \textbf{chronic pain}~\cite{nicholas2019iasp} is pain that persists or recurs for longer than three months. Chronic pain encompasses both chronic primary pain, in which pain itself constitutes the condition (e.g., fibromyalgia), and chronic secondary pain, in which pain is associated with an underlying condition (e.g., cancer, arthritis, nerve injury, post-surgical complications).

Chronic pain is one of the most prevalent and costly health conditions worldwide, affecting between one-fifth and one-half of populations across different countries and generating substantial economic burdens ~\cite{van2023pain}. It has a substantial impact on work participation, contributing to both absenteeism (missing work because of health problems) and presenteeism (reduced productivity while at work despite illness). It is associated with impaired work performance, activity limitations, reduced employment, disability, early retirement, and considerable productivity losses, which are further exacerbated by workplace barriers such as stigma, inadequate organizational support, and excessive workloads~\cite{kawai2017adverse, patel2012impact, blake2025employees, mcdonald2011musculoskeletal}. Beyond its sensory component, chronic pain is consistently associated with impairments in attention, memory, processing speed, and executive functions, thus having broader cognitive consequences~\cite{higgins2018relationship,zhang2021evidence,moore2012interruptive}.  Once viewed primarily as a symptom, chronic pain is now recognized as a distinct condition characterized by intertwined sensory, emotional, cognitive, and social dimensions~\cite{raja2020revised,nicholas2019iasp}. The IASP further emphasizes that pain is a personal experience shaped by biological, psychological, and social factors.

\subsection{The Biopsychosocial Model of Chronic Pain}

Early models of pain, such as the gate control theory~\cite{melzack1965pain} helped establish that pain is not a simple, fixed signal transmitted from the body to the brain. The biopsychosocial approach to chronic pain~\cite{gatchel2007biopsychosocial} offers a model that follows the philosophy of bringing alternative perspectives to the traditional biomedical view, which treats pain purely as a direct signal of tissue damage. This model describes chronic pain as the product of interacting biological, psychological, and social factors, rather than a simple consequence of physical injury. 

A key distinction in this model is between the physical injury or pathology itself and a person's subjective experience of it. Two people with similar physical findings can report very different pain levels and functional impact, depending on psychological and contextual factors~\cite{van2023pain,gatchel2007biopsychosocial}.

Pain is actively shaped and can be amplified or dampened by attention, emotional state, prior experience, and context. Building on this, current research views chronic pain as something that can perpetuate itself through feedback loops between physical sensations, thoughts, emotions, and behavior~\cite{gatchel2007biopsychosocial}. For instance, fear of worsening an injury can lead someone to avoid an activity altogether, which over time reduces physical capacity and increases psychological distress, further reinforcing the pain experience.

This view aligns with the social model of disability~\cite{oliver2013social}, which locates disability not in impairment itself but in its interaction with disabling barriers in society, which is also a principle covered in the UN’s Convention on the Rights of Persons with Disabilities
~\cite{un2006convention}. Just as pain is shaped by context, whether chronic pain becomes disabling depends partly on the barriers and accommodations present in a person's environment, including the workplace.

\subsection{Software Professionals' Well-being}
The Software Engineering literature has largely approached developers' well-being through the lens of mental health~\cite{martinez2026factors,arriel2026understanding,wong2023mental,graziotin2018happens}. However, recent work in the SE community emphasizes that well-being extends beyond psychological states to include physical, cognitive, emotional, and social dimensions~\cite{godliauskas2025well}. This broader perspective highlights how little attention chronic pain has received in software engineering research.

Software development often involves prolonged computer use and sedentary postures. For instance, a comparison of computer office workers with and without reported musculoskeletal pain found that those reporting pain showed significantly worse ergonomic indexes, particularly for chair setup and upper limb postural risk, even though no differences emerged in work-related psychosocial factors between groups~\cite{rodrigues2017differences}. Recent human-computer interaction work focused on \emph{pacing} as an undersupported coping strategy that involves planning the person’s activity and rest to avoid pain flare-ups ~\cite{jamwal2026pacing}. 
Recent research has examined postural differences between software developers with and without chronic neck pain~\cite{kovac2025difference}. Other chronic pain conditions, such as fibromyalgia or migraine, are underexplored in software engineering, despite broader fibromyalgia research being heavily focused on its effects on employment and work capacity in general working populations~\cite{mukhida2020does,lev2022work}. More broadly, a scoping review of chronic pain in episodic illnesses (fibromyalgia, breast cancer, multiple sclerosis, and HIV) found that work-related strategies and barriers have been examined mostly at the individual level, with comparatively little attention paid to workplace and policy-level accommodations that could support sustained work participation~\cite{antao2013chronic}. Overall, software developers have rarely been examined as a distinct population in the chronic pain literature, which has instead focused on computer users broadly, on single diagnoses in isolation, or on general working populations rather than a specific occupational group.

This omission is relevant for software engineers because developer experience (DX) research already treats \textbf{cognitive} and \textbf{affective} states as its central dimensions, along with conation~\cite{fagerholm2012developer,greiler2022actionable}. Cognitive dimensions of DX encompass attention, focus, and mental workload during programming tasks, while affective dimensions capture emotional responses such as frustration, anxiety, and satisfaction that arise from interacting with tools, tasks, and the broader work environment~\cite{fagerholm2012developer}. As noted above, chronic pain is associated with impairments in attention, memory, and executive function, as well as with negative affective states such as anxiety and emotional distress~\cite{higgins2018relationship,zhang2021evidence,moore2012interruptive}. Despite this direct conceptual overlap, the DX literature has not considered chronic pain as a determinant of cognitive or affective states, leaving a population whose daily work is built around sustained, fine-motor, cognitively demanding computer use absent from a body of research that otherwise treats cognition and affect as core outcomes of interest.

\section{Research Design}

We designed an exploratory study to understand how chronic pain shapes software professionals' experiences at work. Our data collection comprised two steps. First, participants completed an online questionnaire covering demographic, professional, and pain-related information to characterize the sample and provide context for the interviews. We then conducted semi-structured interviews exploring how pain interacts with software development activities, how participants manage symptoms while working, and the consequences for sustained participation in the profession. We analyzed the interviews qualitatively to develop the conceptual framework. Questionnaire responses helped contextualize participants' experiences and supported an exploratory analysis of associations between pain characteristics and the categories identified in the interviews.

The following subsections describe participant recruitment, the pre-interview questionnaire, participant demographics, interview procedures, data analysis, member-checking, researcher positionality, and ethical considerations.

\subsection{Participant Recruitment}

We recruited software professionals who had experience in roles related to software development and self-identified as living with chronic pain. Recruitment followed four strategies: (i) social media advertising, (ii) direct outreach through professional networks, (iii) physical posters at our universities, and (iv) snowball sampling. We advertised the study on LinkedIn and Mastodon, contacted potential participants through our professional networks, and invited participants to forward the call to other eligible software professionals. Several participants reported learning about the study through developer chats or community channels, suggesting that the call also circulated informally within software development communities.

In total, 62 software professionals expressed interest. We contacted them by email to arrange interviews, and 34 scheduled and completed an interview. We concluded data collection when later interviews no longer changed the framework. P17's interview was the last to introduce a new framework category or relationship; subsequent interviews allowed us to examine variation across roles, pain conditions, work arrangements, and locations. Our assessment of sample adequacy therefore considered both framework stability and diversity within the sample we could recruit. Recruitment also involved practical constraints, including the time and energy that an interview requires from people living with chronic pain.

\subsection{Pre-interview Questionnaire}

Participants completed an online questionnaire covering demographic, professional, and pain-related information. Alongside characteristics such as professional role and work arrangement, we asked about pain duration, formal diagnosis, current use of pain medication, and interference with daily activities on a 0--10 scale. We developed the questionnaire with the neuropsychology researcher on our team, whose expertise in pain and cognition helped align the pain-related questions with the study's goals. These responses informed our description of the sample and contextualized the interviews.

Because chronic pain assessment requires capturing functional, psychological, and affective dimensions beyond mere severity, we use the Short-Form McGill Pain Questionnaire (SF-MPQ)~\cite{melzack1987short} for its multidimensional coverage of these constructs.
The original MPQ was developed to assess pain as a multidimensional experience, including sensory, affective, and evaluative qualities, since pain is subjective and shaped not only by sensory intensity, but also by emotional, attentional, and contextual factors. We used the short form as it reduces participant burden while preserving structured information about pain quality and intensity.

The SF-MPQ asks participants to rate 15 pain descriptors: 11 sensory descriptors and 4 affective descriptors~\cite{melzack1987short}. Each descriptor is rated from 0 to 3, where 0 means ``none,'' 1 ``mild,'' 2 ``moderate,'' and 3 ``severe.'' The sum of these ratings produces three scores: the Sensory Pain Rating Index (PRI-S), the Affective Pain Rating Index (PRI-A), and the Total Pain Rating Index (PRI-T). The SF-MPQ includes the Present Pain Intensity (PPI) index, where participants select the word that best describes their current pain intensity (e.g., mild, discomforting, distressing, horrible, or excruciating). We also asked participants to report their current pain intensity using the Visual Analog Scale (VAS)~\cite{hawker2011measures}, where 0 indicated no pain and 10 indicated the worst pain imaginable.

\subsection{Participant Demographics}

\begin{table}[t]
\centering
\begin{threeparttable}
\caption{Participant demographic attributes and pain profile (N=34).}
\label{tab:participants}
\scriptsize
\begin{tabular}{p{0.22\linewidth}p{0.66\linewidth}}
\toprule
\multicolumn{2}{l}{\textbf{Demographic attributes}} \\
\midrule
\textbf{Gender} & Male (16); Female (13); Non-binary (4); Prefer not to say (1) \\
\cmidrule(lr){2-2}
\textbf{Age} & 18--24 (1); 25--34 (16); 35--44 (14); 45--54 (2); Over 55 (1) \\
\cmidrule(lr){2-2}
\textbf{Role}\tnote{a} & Software developer (20); Research \& academia (6); UI/UX designer (3); Other technical roles (5) \\
\cmidrule(lr){2-2}
\textbf{Continent} & Europe (12); South America (11); North America (8); Asia (3) \\
\cmidrule(lr){2-2}
\textbf{Work mode} & Remote (21); Hybrid (9); In-person (4) \\
\cmidrule(lr){2-2}
\textbf{Coding experience} & $<$5 years (4); 5--9 years (11); 10--14 years (13); 20--24 years (4); Not reported (2) \\
\midrule
\multicolumn{2}{l}{\textbf{Pain profile}} \\
\midrule
\textbf{Formal diagnosis} & Yes (29); No (4); Prefer not to say (1) \\
\cmidrule(lr){2-2}
\textbf{Pain duration} & More than 3 years (21); 1--3 years (12); 6--12 months (1) \\
\cmidrule(lr){2-2}
\textbf{Pain medication} & Yes (18); No (16) \\
\cmidrule(lr){2-2}
\textbf{Pain interference} & Mean = 5.7, Standard deviation = 2.6 \\
\cmidrule(lr){2-2}
\textbf{PPI} & Mild (3); Discomforting (18); Distressing (11); Horrible (1); Excruciating (1) \\
\cmidrule(lr){2-2}
\textbf{VAS} & Mean = 4.8, Standard deviation = 1.9 \\
\cmidrule(lr){2-2}
\textbf{PRI-S} & Mean = 10.1, Standard deviation = 5.0 \\
\cmidrule(lr){2-2}
\textbf{PRI-A} & Mean = 4.0, Standard deviation = 2.9 \\
\cmidrule(lr){2-2}
\textbf{PRI-T} & Mean = 14.1, Standard deviation = 7.3 \\
\bottomrule
\end{tabular}

\begin{tablenotes}[flushleft]
\footnotesize
\item[a] Three participants were not employed at the time of the interview; their most recent professional role is reported.
\end{tablenotes}
\end{threeparttable}
\end{table}

We interviewed 34 software professionals (P1--P34). Table~\ref{tab:participants} summarizes their demographic attributes and pain profiles. Participants varied in gender, age, geographical region, technical role, and professional experience. Most were between 25 and 44 years old (88\%), worked remotely (62\%), and held software developer roles (59\%). Three participants were not employed at the time of the interview; the table reports their most recent professional roles.
Most participants reported a formal diagnosis (85\%) and had lived with chronic pain for more than three years (62\%), while just over half used pain medication (53\%). Pain interference with daily activities averaged 5.7/10 (SD = 2.6). Participants most commonly described their present pain as discomforting (53\%) or distressing (32\%). Mean scores were 4.8 (SD = 1.9) on the VAS, 10.1 (SD = 5.0) for PRI-S, 4.0 (SD = 2.9) for PRI-A, and 14.1 (SD = 7.3) for PRI-T.

Participants reported a broad range of chronic pain conditions, often with multiple co-occurring diagnoses. Musculoskeletal and spinal conditions were most common (68\%), followed by repetitive strain, tendon, or nerve-compression conditions (29\%) and headache or migraine disorders (29\%). Fibromyalgia was reported by 15\% of participants. Other reported conditions included Ehlers-Danlos syndrome, endometriosis, and irritable bowel syndrome.

\subsection{Semi-Structured Interviews}

We conducted semi-structured interviews to explore how participants experience and navigate chronic pain at work. Semi-structured interviews are well-suited for empirical software engineering because they combine a consistent interview structure with the flexibility to explore unexpected topics and follow emerging themes during the conversation~\cite{hove2005experiences}. All interviews were conducted remotely via video conferencing by the first author. Before each interview, participants reviewed the informed consent form, agreed to audio and video recording, and received a brief explanation of the study objectives, interview procedure, and their rights as participants. The interviews lasted on average 35 minutes. We conducted 11 interviews in Brazilian Portuguese and 23 in English. The first author translated the Portuguese transcripts into English for analysis.

The interview protocol was organized around few broad themes. We began by establishing rapport and asking participants about their chronic pain history and professional background. We then explored the relationship between work arrangements (e.g., remote, hybrid, and in-person work) and pain management. The core of the interview focused on how chronic pain interacted with software development activities as well as which aspects of the work aggravated participants' symptoms. We also discussed the strategies participants used to manage pain while working. Finally, we explored workplace accommodations, organizational support, and participants' recommendations for improving the experience of software professionals living with chronic pain. During interviews, we used follow-up questions to clarify responses and investigate emerging topics, consistent with recommended practices for semi-structured interviewing in software engineering~\cite{hove2005experiences}. The complete interview protocol is included in our supplementary materials.

\subsection{Data Analysis}

We qualitatively analyzed the interview transcripts through several rounds of open and axial coding~\cite{strauss1998basics}. We focused on how participants experienced chronic pain at work, which aspects of work worsened their symptoms, and how they managed pain while continuing to work. During open coding, the first author assigned initial codes to excerpts from interviews. In regular team meetings, we reviewed these codes, merged overlapping codes, split broad codes, and refined them into categories with agreed definitions~\cite{patton2014qualitative}. We grouped related \textbf{\textit{categories}} (like sitting still for long hours and deadline pressure) into broader \textbf{concepts} (e.g., work demands) that formed the main parts of the framework.

During axial coding, we examined how participants described relationships between the main concepts. For example, P13 described becoming absorbed in a task and delaying pain-relief cream until discomfort increased. This helped us identify the relationship between \textbf{work demands} and \textbf{pain aggravation}. The whole team, including a neuropsychologist, discussed and refined the concepts and their relationships. We identified \textbf{contextual conditions}, such as workplace flexibility and personal resources, as crosscutting influences that could affect several concepts and relationships across the framework. These relationships reflect participants' descriptions and do not establish cause and effect or a sequence that applies to everyone.

Throughout the analysis, we compared excerpts across interviews to identify similarities and differences in participants' experiences~\cite{glaser2017discovery}. We discussed these comparisons as a team and agreed on revisions to the categories, definitions, and framework~\cite{seaman2025qualitative,lenberg2024qualitative}. The final codebook contains 42 categories organized under seven concepts, each category with a definition, inclusion and exclusion criteria, participant counts, and representative quotations.

To complement the qualitative analysis, we explored associations between four pain-profile measures (PPI, PRI-S, PRI-A, and pain interference) and 33 framework categories, yielding 132 comparisons. The categories covered \textbf{work demands}, \textbf{chronic pain experience}, \textbf{impacts on work}, \textbf{pain management}, and \textbf{sustained participation}. Pain aggravation and contextual conditions were excluded from this analysis. For each participant, categories were coded as 1 if mentioned in their interview and 0 otherwise. We calculated Spearman's rank correlations and obtained two-sided p-values using permutation tests. No correction for multiple comparisons was applied. We interpreted these associations cautiously as exploratory findings that could include chance results, rather than as evidence validating the framework or establishing cause and effect.

\subsection{Member-checking}
We emailed all participants the framework and a brief description, asking what reflected their experiences, what was missing or overemphasized, and whether any relationships seemed incorrect. The exact email is included in our supplementary materials. Responding was optional, recognizing the time and energy required from people living with chronic pain. Six participants replied, indicating that the framework reflected their experiences. One participant explicitly suggested making personal coping styles, including \textit{workaholism}, more explicit; we incorporated this suggestion by broadening the terminology used for contextual conditions.

\subsection{Researcher Positionality}

This study was conducted by a multidisciplinary team of three software engineering researchers and one neuropsychologist with expertise in pain and cognition. Consistent with recent discussions on researcher reflexivity and positionality in Empirical Software Engineering~\cite{seaman2025qualitative,felix2025,lenberg2024qualitative}, we acknowledge that our perspectives on chronic pain are informed not only by disciplinary knowledge but also by personal experiences and relationships. One of the lead authors lives with fibromyalgia, bringing a direct perspective on the often invisible effects of chronic pain on cognition, energy levels, and professional activities. Another author has a close family member who experiences chronic pain, bringing a relational perspective on its emotional and practical consequences.

Recognizing our proximity to the topic, we adopted a reflexive stance throughout the research process. We engaged in regular interdisciplinary discussions to examine how our experiences and disciplinary backgrounds might influence the study design, data interpretation, and presentation of findings. The neuropsychologist collaborator brings expertise in pain processing and its cognitive effects from both clinical and research perspectives. This expertise helped relate the findings on cognition, fatigue, attention, and emotion to established pain and cognitive science literature. During analysis, we continuously reflected on whether emerging themes were grounded in participants' perspectives or shaped by our own assumptions, revising interpretations when necessary. To strengthen the credibility of our findings, we combined multiple data sources, engaged in team-based analysis discussions, and conducted member-checking. Rather than trying to remove subjectivity, we acknowledge it and manage its influence throughout the study.

\subsection{Ethics and Supplementary Material}

The Ethics Committee of the Faculty of Science of Radboud University approved this study. In accordance with the relevant guidelines and regulations, we report participant profiles in aggregate and use participant identifiers for quotations and examples, with personally identifiable information removed. To support transparency and reproducibility, we include aggregated, non-identifiable study materials and analysis artifacts in our supplementary package.\footnote{\url{https://doi.org/10.6084/m9.figshare.32851850}}

\section{Findings}

\begin{figure}[!t]
    \centering
    \includegraphics[width=\columnwidth]{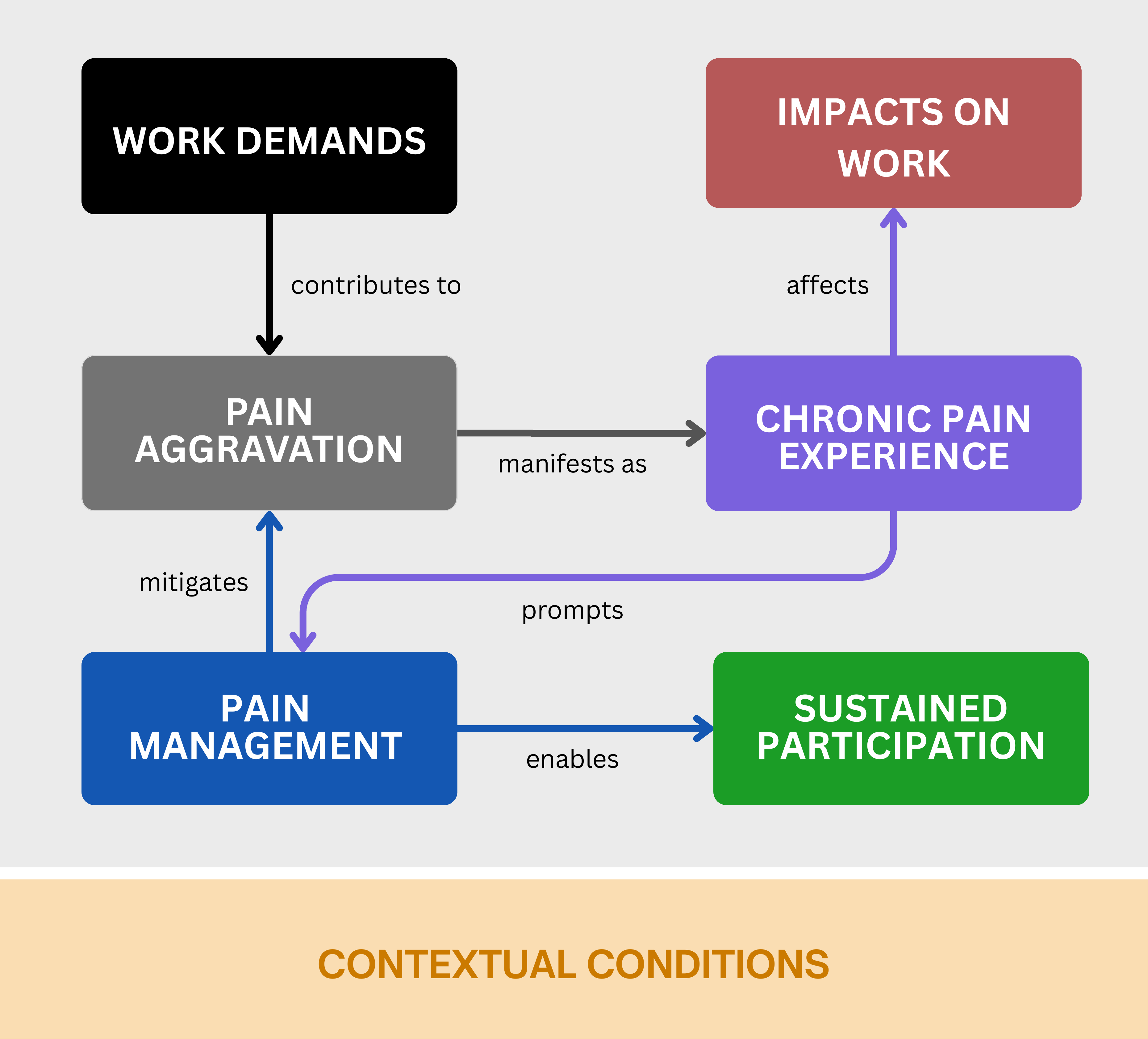}
    \caption{Main concepts and relationships of our framework.}
    \label{fig:framework}
\end{figure}

\begin{figure*}[!t]
    \centering
    \includegraphics[width=\linewidth]{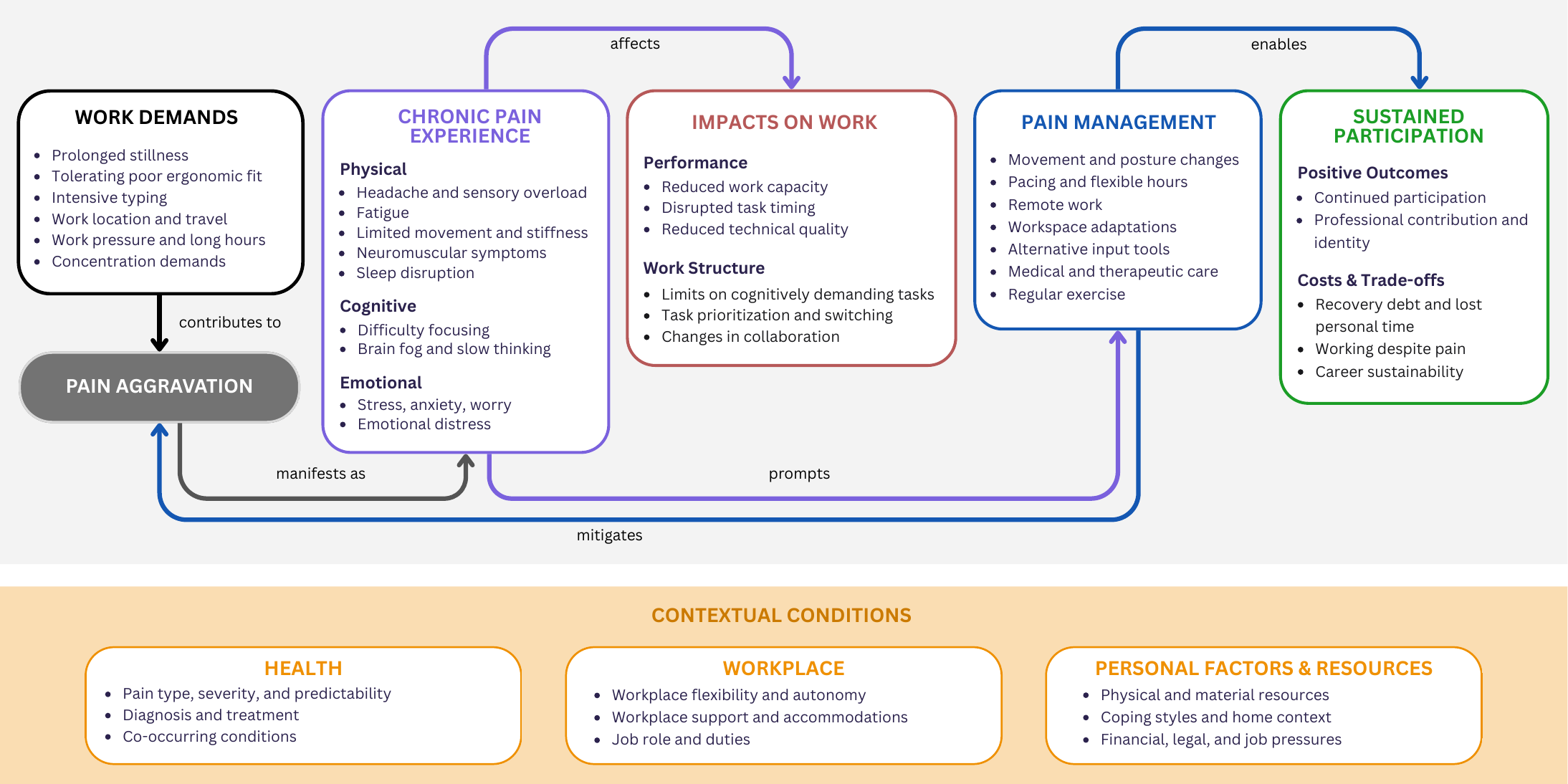}
    \caption{Categories within each concept of the framework, including contextual conditions grouped into health, workplace, and personal factors and resources.}
    \label{fig:framework-detailed}
\end{figure*}

From our qualitative analysis of the interviews, we derived a framework describing how software professionals experience and manage chronic pain at work. Figure~\ref{fig:framework} summarizes the main concepts (as boxes) and their relationships (as arrows). Figure~\ref{fig:framework-detailed} shows the full framework with the categories within each concept. In the following subsections, we indicate main concepts in \textbf{bold} font, relationships in \textit{italics}, and categories in \textit{\textbf{bold + italics}}. Numbers in parentheses indicate how many participants mentioned each category during their interview.

Participants described several \textbf{work demands} as \textit{contributing to} \textbf{pain aggravation}. For participants already living with chronic pain, this aggravation \textit{manifests as} worsening physical, cognitive, and emotional symptoms within their \textbf{chronic pain experience}. Participants described how these experiences \textit{affected} their work (\textbf{impacts on work}), including their work capacity, task completion, perceived technical quality, task selection, and collaboration. These experiences also \textit{prompted} \textbf{pain management} through adaptations to routines, tools, workspaces, and schedules. Participants described these strategies as \textit{mitigating} pain aggravation and \textit{enabling} \textbf{sustained participation} in the workplace. \textbf{Contextual conditions}, including their health, workplace, and home life, emerged as a crosscutting effect that affected the demands they faced, how they managed pain, and what it took to keep working.

\subsection{Work Demands and Pain Aggravation}

Participants described \textbf{work demands} that made existing pain worse, triggered flare-ups, or made symptoms harder to manage (\textbf{pain aggravation}). Here, flare-ups refer to temporary periods when pain or related symptoms become worse than usual. The \textit{contributes to} relationship reflects these descriptions. 

The most frequently mentioned demands concerned posture and equipment. Participants described how \textit{\textbf{prolonged stillness}} (18) during programming, debugging, meetings, and other computer activities could worsen pain. P17, for example, described pain starting after several hours of sitting. Participants also described worsening symptoms when \textit{\textbf{tolerating poor ergonomic fit}} (17), such as working with desks, chairs, monitors, or input devices that did not suit their bodies or needs. P3 described worsening pain after changing their desk and chair. Others described symptoms associated with particular keyboards or trackpads.

Participants also described how \textit{\textbf{intensive typing}} (14), including sustained keyboard and mouse use, repeated shortcuts, IDE navigation, and long programming sessions, could worsen their symptoms. P4 said, ``\textit{If I keep typing for a prolonged period, I feel more discomfort.}'' P14 similarly described wrist pain after coding for several hours.

Different work arrangements (like working from home or traveling) could affect pain through differences in equipment and opportunities to rest. These effects depended on \textbf{contextual conditions}, discussed later in Section~\ref{sec:context}. Remote work, for example, could support pain management by offering greater control over the environment and more opportunities to rest (Section~\ref{sec:pain_management}). However, participants also described how \textit{\textbf{work location and travel}} (14) could worsen symptoms. P5 described spending more time sitting without moving after switching to working from home. P9 described an inadequate home workstation, while P15 associated unfamiliar offices and hotel stays with flare-ups.

Participants described \textit{\textbf{work pressure and long hours}} (7) in their software development jobs, including deadlines, urgent requests, heavy workloads, and sprint pressure. These demands could increase stress and worsen symptoms. P2 described how urgent requests brought on neck pain, saying, ``\textit{Urgent demands arrive [and] I feel under pressure, then I already get that pain in my neck.}'' P30 described finger pain getting worse after working 14 hours a day for three days on the same task.

Participants described needing to concentrate for long periods during development tasks. These \textit{\textbf{concentration demands}} (6) could make them less aware of their posture, symptoms, or pain management needs while absorbed in a task. P14 described noticing an uncomfortable posture only after hours had passed. P13 became absorbed in a problem and forgot to apply pain-relief cream.

\subsection{Chronic Pain Experience}

The \textbf{chronic pain experience} describes what living and working with pain feels like physically, cognitively, and emotionally. It can involve bodily discomfort, difficulty thinking or focusing, and feelings such as frustration or worry.

Participants experienced \textit{\textbf{limited movement and stiffness}} (16), including difficulty turning their necks, lifting their arms, bending their wrists, or changing position. Some also experienced \textit{\textbf{headache and sensory overload}} (10), including migraines and sensitivity to light, sound, or smell. Participants described \textit{\textbf{neuromuscular symptoms}} (10) as tingling, numbness, weakness, reduced sensitivity, and sensations like electric shocks.

Pain also affected sleep and energy. Participants described \textit{\textbf{sleep disruption}} (9), including difficulty falling asleep or finding a comfortable position. P2 described being unable to relax enough to sleep. Others experienced persistent \textit{\textbf{fatigue}} (8), feeling tired or physically exhausted. P1 described tiredness throughout the day and feeling as though their body was not functioning.

Cognitive experiences included \textit{\textbf{difficulty focusing}} (17), as pain and bodily sensations repeatedly drew participants' attention away from what they were doing. P13 said, ``\textit{If I would have spikes of pain while I'm trying to focus, that was really hard to maintain focus.}'' Participants also experienced \textit{\textbf{brain fog and slow thinking}} (7), describing unclear, foggy, or slower thought. P2 described being unable to think clearly because of pain.

Participants experienced \textit{\textbf{emotional distress}} (9), including frustration, sadness, and helplessness. P2 described feeling distressed over several months while trying to understand their condition. Participants also described \textit{\textbf{stress, anxiety, worry}} (6) about their pain and what might happen next. This included fear of further pain and worry about how they would feel during the next workday or an upcoming deadline.

\subsection{Impacts on Work}

Pain may change how a workday unfolds, from the time needed to finish a task to the choice of which tasks to attempt. These \textbf{impacts on work} concerned what participants could accomplish and how well they could do it (performance), and how they organized tasks and worked with others (work structure), as shown in Figure~\ref{fig:framework-detailed}.

Participants described \textit{\textbf{reduced work capacity}} (19) that limited how long they could work and sometimes made them stop for the day. P17 described being unable to work an eight-hour day with only a lunch break. P22 said, ``\textit{Sometimes maybe I'll only be able to work for four hours and then it will be so painful that I just stop.}''

Participants also experienced \textit{\textbf{disrupted task timing}} (6), with tasks taking longer, being interrupted or postponed, or becoming harder to plan. P25 described completing a task in one hour when feeling well but being unable to finish it after three hours when in pain. P17 described customer incidents taking longer to process, while P24 described stopping a task and submitting it the next day.

Participants reported \textit{\textbf{reduced technical quality}} (6) due to the mistakes they made, the errors they missed during debugging, and how ready they felt their work was for review. P13 reported making ``\textit{much more mistakes.}'' P23 described difficulty spotting an error while debugging code, even though it was right in front of them. They spent an hour and a half before noticing it. P29 described producing work that still needed further development before they would ask someone to review it, saying, ``\textit{It's just a very rough draft. It's not to the level that I would share with someone for a review.}'' These findings are based on what participants reported about their work; we did not independently evaluate their code.

Pain also affected which development activities participants could manage, particularly those requiring sustained attention and reasoning. Participants described \textit{\textbf{limits on cognitively demanding tasks}} (10) when writing test cases, preparing documentation, designing, responding to reviewer comments, and performing security checks. P9 explained that writing test cases required time to stop, reflect, and think carefully, which became particularly difficult during a pain flare-up. P15 said, ``\textit{software engineering is so cognitively demanding. If I can't focus, I can't get things done.}'' P32 similarly described struggling with problem-solving and logic tasks. Participants also described \textit{\textbf{task prioritization and switching}} (12), adjusting which activities they undertook and in which order. Some switched from typing-intensive coding to reading or outlining documentation, while others postponed demanding tasks or had work reassigned.

Participants described \textit{\textbf{changes in collaboration}} (10), adjusting how and when they joined meetings or took part in conversations. Some stood during meetings, avoided long conversations, or changed communication channels. P17 described attending some meetings from bed with the camera off. P21 prioritized important meetings and skipped others.

\subsection{Pain Management}
\label{sec:pain_management}

Managing pain was an ongoing part of participants' working lives. Their symptoms \textit{prompted} \textbf{pain management} through treatment and changes to how, when, and where they worked. Participants often combined several strategies, which they described as a way to \textit{mitigate} \textbf{pain aggravation}.

Participants used \textit{\textbf{medical and therapeutic care}} (22), including medication, topical treatment, injections, physiotherapy, occupational therapy, and massage. During the workday, participants made \textit{\textbf{movement and posture changes}} (21), such as standing, stretching, walking, lying down, or changing position. P1 described stretching during meetings, while P15 alternated between sitting and standing. Participants also used \textit{\textbf{regular exercise}} (18), including Pilates, yoga, strengthening, walking, and climbing, to manage symptoms over time.

Participants adjusted when and how long they worked through \textit{\textbf{pacing and flexible hours}} (20). They planned breaks, used Pomodoro routines, logged off early, took sick days, or paused work before returning later. P17 described resting for several hours in the afternoon and working later when flexible hours allowed it. Participants also made \textit{\textbf{workspace adaptations}} (19), adjusting chairs, desks, monitors, and headsets, or using heating pads and other supports to make working more comfortable.

Working from home could make these strategies easier to use. Participants described how \textit{\textbf{remote work}} (16) allowed them to lie down, move privately, control lighting and noise, and avoid commuting. P22 described being able to rest, sleep, or walk before returning to work. This helps explain why remote work appeared both among work demands and among pain management strategies. An inadequate home setup could worsen symptoms, while control over the environment and schedule could help participants manage them.

When typing or mouse use became painful, participants used \textit{\textbf{alternative input tools}} (9) to change how they interacted with computers. These included ergonomic keyboards and mice, dictation, voice-controlled programming, and eye tracking. Two participants described using Talon,\footnote{\url{https://talonvoice.com}} a system for voice-controlled programming and computer use. P18 said, ``\textit{I thought of ways to control my computer by voice [and I] came across Talon. [...] I can get by using voice.}'' These tools offered other ways to write code and use development tools when standard keyboard or mouse input was challenging.

\subsection{Sustained Participation and Its Costs}

Pain management helped participants keep working, \textit{enabling} \textbf{sustained participation}. However, continuing to work could involve ongoing pain, more recovery time, and less personal time. Our framework separates positive outcomes from costs and trade-offs.

A positive outcome was \textit{\textbf{continued participation}} (14), with participants remaining productive, returning after rest, or staying in software development. P15 described treatment as making continued work manageable. Remaining engaged supported participants' \textit{\textbf{professional contribution and identity}} (8), helping them remain useful, reliable, and involved in their teams. P7 described trying to contribute without slowing the team's progress, while P15 described organizational support as encouraging loyalty and longer-term contribution.

Continuing to work could involve \textit{\textbf{recovery debt and lost personal time}} (16). These costs included an accumulated need for rest when recovery was postponed, and personal time spent completing work or recovering afterward. P1 described working longer to compensate for perceived lost output, followed by severe back pain. Lost personal time could also result from extending the workday to accommodate breaks. P17 said, ``\textit{I end up working until like 7:30 to 8:00 PM because I took hours long breaks.}'' Flexible hours could therefore help participants continue contributing while taking time away from their personal lives.

For some participants, \textit{\textbf{working despite pain}} (11) remained part of everyday life. Symptoms were still present but manageable enough to continue development activities. P22 described living and working this way for many years. Pain management made work possible without necessarily removing the pain.

Concerns about \textit{\textbf{career sustainability}} (10) included limited career growth, job loss, and uncertainty about how long participants could remain in software development. P5 described fewer opportunities for career growth, P23 reported dismissal connected to disability, and P10 questioned how they could continue working for another 40 years.

\subsection{Contextual Conditions}
\label{sec:context}

Managing pain and continuing to work also depended on participants' circumstances. These \textbf{contextual conditions} shaped the demands they faced, the strategies they could use, and the costs of staying in work.

Participants differed in their \textit{\textbf{pain type, severity, and predictability}} (18). Some experienced constant background pain, while others had periods of more intense or unpredictable symptoms. Their experiences with \textit{\textbf{diagnosis and treatment}} (16) included uncertainty about their condition, access to care, available treatments, and side effects. Participants had to manage \textit{\textbf{co-occurring conditions}} (11), such as migraine, sleep disorders, ADHD, autism, and mental-health conditions, alongside chronic pain. These health circumstances affected both how they experienced pain and their options for managing it.

The workplace could make adjustments easier or harder to arrange. Participants described how \textit{\textbf{workplace support and accommodations}} (17) depended on managerial trust, supportive colleagues, HR processes, and access to equipment. \textit{\textbf{Workplace flexibility and autonomy}} (14) gave participants control over their hours, location, breaks, and opportunities to rest. Their \textit{\textbf{job role and duties}} (16) shaped how much time they spent coding, attending meetings, managing others, traveling, or working with clients. These differences affected the demands participants faced and how much they could adjust their work.

Personal factors and resources shaped which pain management strategies were practical. Access to \textit{\textbf{physical and material resources}} (16) included suitable home workspaces, ergonomic equipment, adaptive technologies, and affordable tools. Participants' \textit{\textbf{coping styles and home context}} (11) included how they handled work and stress, caregiving, household responsibilities, living arrangements, and social support. For example, resting during the workday depended partly on having a suitable space and enough support at home. Their options could be further limited by \textit{\textbf{financial, legal, and job pressures}} (14), which affected whether they could reduce their hours, change jobs, disclose their needs, or stop working when pain worsened. Such circumstances help explain why participants with similar symptoms could have different options for managing pain and different costs of continuing to work.

\subsection{Exploratory Associations with Pain Profiles}

We explored associations between participants' pain profiles and the categories mentioned in their interviews (see supplementary materials for the complete results). Positive correlations indicate that participants who mentioned a category tended to have higher pain scores, while negative correlations indicate that they tended to have lower scores. These comparisons concern whether a category was mentioned, not how often or how strongly participants experienced it.

Higher affective pain (PRI-A) was positively associated with mentions of \textit{\textbf{emotional distress}} ($\rho=0.42$, $p=0.016$) and \textit{\textbf{work pressure and long hours}} ($\rho=0.35$, $p=0.046$), and negatively associated with mentions of \textit{\textbf{task prioritization and switching}} ($\rho=-0.36$, $p=0.038$) and \textit{\textbf{working despite pain}} ($\rho=-0.53$, $p=0.002$). Higher sensory pain (PRI-S) was positively associated with mentions of \textit{\textbf{work pressure and long hours}} ($\rho=0.34$, $p=0.048$), and negatively associated with mentions of \textit{\textbf{working despite pain}} ($\rho=-0.40$, $p=0.021$). Higher present pain intensity (PPI) was negatively associated with mentions of \textit{\textbf{sleep disruption}} ($\rho=-0.39$, $p=0.037$).

These associations do not establish cause and effect. Participants who did not mention a category may still have experienced it. The results use unadjusted permutation p-values across 132 comparisons, so some associations may have appeared by chance. They should therefore be interpreted cautiously as exploratory findings rather than evidence supporting the framework's relationships.

\section{Discussion}

\subsection{Comparison to the Biopsychosocial Model of Pain}

The biopsychosocial model describes chronic pain as the result of interacting biological, psychological, and social factors rather than solely tissue damage~\cite{gatchel2007biopsychosocial}. Work is recognized as one social factor, but the model says little about how the demands of a particular profession shape pain experiences. Our framework instantiates these general mechanisms within software development, tracing how \textbf{work demands} contribute to \textbf{pain aggravation}, which manifests as worsening symptoms within participants' existing \textbf{chronic pain experience}, reshapes software development activities (\textbf{impacts on work}), and prompts \textbf{pain management} that supports \textbf{sustained participation}. Our main contribution lies in showing how these relationships appear in software development activities.

The clearest example concerns attention. Participants described how \textit{\textbf{concentration demands}} delayed awareness of pain during programming and debugging until symptoms became difficult to ignore, after which \textit{\textbf{difficulty focusing}} became one of the most frequently reported pain experiences. This suggests an occupation-specific interaction between the attentional demands of software development and pain regulation. Participants adapted both their work practices and their development tools. Their use of \textit{\textbf{alternative input tools}}, such as voice-controlled programming (e.g., Talon), highlights accessibility needs specific to software development. Likewise, \textit{\textbf{task prioritization and switching}} reflected the physical, cognitive, and collaborative demands of development rather than simply reducing workload.

Other aspects of our framework are broadly consistent with the biopsychosocial literature. In particular, \textit{\textbf{pacing and flexible hours}} and \textit{\textbf{workplace flexibility and autonomy}} align with previous work linking perceived control to better pain outcomes, although perceived control was not directly measured in our study. Similarly, \textit{\textbf{task prioritization and switching}} may reflect fear of symptom escalation or pragmatic symptom management, but our interview data do not allow us to disentangle these explanations.

\subsection{A Physical Dimension for Developer Experience}

Our findings suggest investigating how physical experiences are represented in developer experience (DX). Existing DX research considers cognitive and affective aspects of developers' work~\cite{fagerholm2012developer,greiler2022actionable}. In our framework, \textbf{chronic pain experience} includes physical, cognitive, and emotional experiences that participants often described together. For example, \textit{\textbf{concentration demands}} could delay attention to symptoms, while pain could lead to \textit{\textbf{difficulty focusing}}. Participants described physical symptoms alongside \textit{\textbf{reduced work capacity}} and perceived \textit{\textbf{reduced technical quality}}. These descriptions suggest that physical experiences are relevant to understanding how software professionals living with chronic pain think, feel, and carry out development activities.
These findings provide a starting point for investigating an explicit physical dimension of DX. Future studies could examine whether this dimension adds to existing models, how it relates to cognitive and affective experiences, and how it could be measured. Such work would need to consider differences across development roles, health conditions, and working environments.

\subsection{Implications for Software Engineering}

\noindent \textbf{Practitioners:}
Participants’ descriptions suggest that proactive adjustments may help them manage pain during work. This can mean \textit{scheduling demanding tasks for lower-pain periods}, reserving lighter tasks for flare-up days, taking breaks early, changing posture during meetings, or using alternative ways to participate when typing or speaking is hard. This echoes \textit{pacing strategies} used in other chronic pain populations~\cite{antao2013chronic} and the finding that decision control over how work is done helps sustain participation~\cite{mukhida2020does}.
Chronic pain is often invisible, as someone can be present in a meeting while managing real pain, fatigue, or brain fog. Teams can \textit{normalize flexible participation} such as standing during meetings, cameras off, asynchronous communication, or switching channels when symptoms worsen. This matters for developer performance because pain affects attention, memory, and executive functions~\cite{higgins2018relationship,zhang2021evidence,moore2012interruptive}, which relate to cognition, a dimension developer experience research already treats as central~\cite{fagerholm2012developer,greiler2022actionable}.
\textit{Alternative input tools}, ergonomic keyboards, vertical mice, and dictation, for some developers, make continued programming possible. Teams should normalize their use in meetings, pairing, and code review.

\noindent \textbf{Companies:}
Accommodations should address health needs, work demands, flexibility, and role expectations alongside the physical setup~\cite{gatchel2007biopsychosocial,oliver2013social,un2006convention}. \textit{Flexibility} should be treated as an accessibility resource. Participants relied on remote work, flexible hours, and task switching to manage pain. Explicit policies could formalize flexible schedules and meeting alternatives, reducing reliance on individual managers' goodwill~\cite{blake2021pain}. \textit{Remote work} can worsen pain through unsuitable home setups and reduced movement, or help through greater control over posture and rest~\cite{de2024hybrid}. Its usefulness therefore depends on individual resources. Managers should receive training on \textit{invisible and fluctuating disability}~\cite{blake2021pain} and recognize how deadlines and travel can limit pain management. Workload planning and career reviews should account for \textit{reduced or fluctuating capacity} to avoid penalizing employees for chronic pain. Reports of \textit{recovery debt} suggest examining formal recovery time after intense work, flare-ups, or travel. Finally, \textit{job insecurity} may discourage disclosure and requests for adjustments, limiting access to support.

\noindent \textbf{Researchers:}
Chronic pain deserves attention in SE well-being research alongside mental health, stress, and burnout~\cite{martinez2026factors,arriel2026understanding,wong2023mental,graziotin2018happens}. Future studies could investigate a physical dimension of DX~\cite{fagerholm2012developer,greiler2022actionable}, differences across development roles, and relationships between pain profiles, development activities, and sustained participation. Diary and longitudinal studies could track changing symptoms and recovery debt.
Tool-oriented research could examine how IDEs, code review tools, and collaboration platforms support alternative input and pacing, building on recent HCI research~\cite{jamwal2026pacing}. Such work should consider autonomy, privacy, and professional identity. Given the importance of work identity for people with chronic illness~\cite{lev2022work}, future studies could examine whether accommodations such as role redesign support professional identity and long-term retention.

\section{Limitations and Threats to Validity}

\noindent \textbf{Internal validity.}
Our framework captures participants' descriptions and does not establish causal relationships. Our backgrounds and experiences may have influenced how we understood and interpreted participants' descriptions. We examined our interpretations through iterative coding, constant comparison, team agreement, and reflection on our backgrounds and lived experiences~\cite{lenberg2024qualitative,seaman2025qualitative}. Member-checking feedback supported the framework's correspondence with participants' experiences; however, only six participants responded.

\noindent \textbf{External validity.}
Our sample is not representative of all software professionals with chronic pain. Open recruitment may have attracted people more motivated to discuss pain or experiencing greater work-related impacts. People whose pain made interviewing difficult may be underrepresented. Although later interviews introduced no new categories or relationships, the framework may not cover all experiences or apply equally across settings.

\noindent \textbf{Construct validity.}
Our findings rely on retrospective self-reports and are subject to recall bias. All participants described past work experiences, including three who were no longer employed for reasons related to pain. Translation from Portuguese to English may have affected nuances of meaning. Participants may have experienced symptoms, faced difficulties, or used strategies that they did not mention during their interview.

\noindent \textbf{Conclusion validity.}
We examined correlations without correcting for multiple testing, so some associations may have appeared by chance and should be interpreted cautiously.

\section{Conclusion}

\textit{\textbf{What does it take for software professionals to keep contributing every day despite their chronic pain condition?}} Our findings suggest that the answer is ongoing pain management and \textit{ongoing adjustments to routines and work practices}, often at a hidden price. Although most participants remained engaged in their professional activities, they often did so by accumulating recovery debt, working through persistent symptoms, and worrying about long-term career sustainability. \textit{\textbf{Their code might run just fine, but they often do not.}}

Through interviews with 34 software professionals living with chronic pain, we developed a conceptual framework summarizing how participants linked work demands to worsening pain and described changes in work practices, pain management, and sustained participation. Situating these experiences within software engineering shows how general mechanisms described in the biopsychosocial model of pain take on occupation-specific forms. We hope this framework provides a foundation for future research in Software Engineering and Developer Experience and contributes to the design of more inclusive and sustainable SE workplaces.

\section*{Acknowledgments}
We thank our participants for sharing their experiences and dedicating some of their precious \emph{spoons}\footnote{\url{https://en.wikipedia.org/wiki/Spoon_theory}} to this research.

\balance
\bibliographystyle{IEEEtran}
\bibliography{main}
\end{document}